\documentclass[letterpaper,11pt]{article} 
\usepackage[letterpaper,hmargin=1in,vmargin=1in]{geometry} 
\usepackage{fullpage}
\usepackage[pdfstartview=FitH,colorlinks,linkcolor=blue,citecolor=blue]{hyperref} 

\usepackage[T1]{fontenc}
\usepackage{amsthm} 
\usepackage{color} 
\usepackage{times} 
\usepackage{microtype,pdfsync} 
\usepackage{amsmath,amsfonts,amssymb} 
\usepackage{url} 
\usepackage{xspace,paralist}
\usepackage{lmodern}
\usepackage{float}
\floatstyle{boxed}
\restylefloat{figure}
\usepackage{tikz}
\usetikzlibrary{shapes.geometric}
\usetikzlibrary{positioning}

\usepackage{algpseudocode}
\usepackage{algorithm}

\theoremstyle{plain} 

\newtheorem{theorem}{Theorem}[section] 
\newtheorem{lemma}[theorem]{Lemma}

\theoremstyle{definition}

\numberwithin{equation}{section}

\newcommand{\bin}{{\{0,1\}}}

\newcommand{\Ds}{{\mathcal{D}}}

\newcommand{\Fs}{{\mathcal{F}}}

\newcommand{\Ps}{{\mathcal{P}}}

\newcommand{\Xs}{{\mathcal{X}}}
\newcommand{\Ys}{{\mathcal{Y}}}

\newcommand{\ignore}[1]{}

\begin{document}

\title{A Note on the Quantum Collision and Set Equality Problems}
\author{{\sc Mark Zhandry} \\
Stanford University, USA \\
{\tt mzhandry@stanford.edu} }

\date{}
\maketitle

\begin{abstract} The results showing a quantum query complexity of $\Theta(N^{1/3})$ for the collision problem do not apply to random functions.  The issues are two-fold.  First, the $\Omega(N^{1/3})$ lower bound only applies when the range is no larger than the domain, which precludes many of the cryptographically interesting applications.  Second, most of the results in the literature only apply to $r$-to-1 functions, which are quite different from random functions.

Understanding the collision problem for random functions is of great importance to cryptography, and we seek to fill the gaps of knowledge for this problem.  To that end, we prove that, as expected, a quantum query complexity of $\Theta(N^{1/3})$ holds for all interesting domain and range sizes.  Our proofs are simple, and combine existing techniques with several novel tricks to obtain the desired results.

Using our techniques, we also give an optimal $\Omega(N^{1/3})$ lower bound for the set equality problem.  This new lower bound can be used to improve the relationship between classical randomized query complexity and quantum query complexity for so-called permutation-symmetric functions.

\end{abstract}

\section{Introduction}

\label{sec:intro}

A \emph{collision} for a function $f$ consists of two distinct inputs $x_1,x_2,x_1\neq x_2$ that map to the same value: $f(x_1)=f(x_2)$.  In this note, we explore the difficulty of computing collisions in the quantum query model: how many quantum queries to an unknown function $f$ are required to produce a collision?  Let $M$ be the size of the domain of $f$, and $N$ be the size of the codomain.

Brassard, H{\o}yer, and Tapp~\cite{BHT1997} give a quantum algorithm (henceforth called the BHT algorithm) requiring $O(M^{1/3})$ quantum queries to any two-to-one function $f$ to produce a collision with overwhelming probability.  Ambainis~\cite{Ambainis2003} gives an $O(M^{2/3})$ algorithm (which we will call Ambainis's algorithm) for finding a collision in an arbitrary function $f$, guaranteed that it contains at least one collision.  This latter problem is related to the so-called element distinctness problem, where one is asked to distincguish between an injective function and a function with a single collision.

On the lower bound side,  most of the prior work has also focused on two-to-one functions.  In the case where the domain and co-domain are the same, Aaronson and Shi~\cite{AS2004} and Ambainis~\cite{Ambainis05} prove an $\Omega(N^{1/3})=\Omega(M^{1/3})$ lower bound for two-to-one functions.  The results also generalize to $r$-to-one functions.  These results also imply an $\Omega(M^{2/3})$ lower bound for the element distinctness problem.  

While the above results provide matching upper and lower bounds for the problems they analyze, they have a couple crucial limitations:
\begin{itemize}
	\item Random functions are, with overwhelming probability, not $r$-to-one functions.  Many images will have only one pre-image, and others will have many pre-images.  Thus, the above results are not directly applicable to random functions. 
	\item The above lower bounds are proved by showing that $\Omega(N^{1/3})$ queries are necessary to distinguish a two-to-one function from a permutation.  Since a collision is proof that a function is not a permutation, this implies an $\Omega(N^{1/3})$ lower bound for finding a collision.  This is the reason the domain and co-domain must be the same.  For random functions, it is perfectly reasonable to discuss the collision problem in settings where the co-domain is much, much smaller than the domain, and the above approaches will therefore not work directly.
\end{itemize}

Yuen~\cite{Yuen2013} overcomes some of the above limitations by proving, in the case that the domain and codomain are the same size, the BHT algorithm does indeed produce a collision for random functions after $O(N^{1/3})$ queries.  For a lower bound, Yuen proves that $\Omega(N^{1/5}/\log N)$ queries are necessary to produce a collision, using a combination of the $r$-to-one lower bounds from \cite{AS2004} and the quantum adversary method.  However, the proof works by proving the indistinguishability of a random function from a random permutation, and thus still requires the domain and codomain to be the same.  

We argue, however, that \emph{conditional} quantum lower bounds for the collision problem for random functions are actually implicit in the earlier works of Boneh et al.~\cite{BDFLSZ11,Zha12a,BZ13b}.  Their results show, under several different computational assumptions, a weak lower bound of $(\log N)^{\omega(1)}$.  Basically, they construct various schemes in the \emph{quantum random oracle model}, in which all parties can make quantum queries to a shared random function.  In many of the schemes, a collision for the random function would break the security of the scheme, and can hence be used to break the underlying computational assumption.  By making \emph{subexponential} hardness assumptions, we can even improve the lower bound to $N^{\Omega(1)}$ queries.

\paragraph{Relevance.}  Why do we care about the quantum query complexity of finding collisions in random functions?  For starters, the quantum collision problem for random functions is a very natural problem, perhaps more natural than two-to-one functions.

Second, when looking at random functions, the collision problem and the element distinctness problem turn out to essentially be one and the same.  In particular, when the codomain size $N$ is quadratic in the domain size $M$, a random function will be injective with constant probability, and will have a collision with constant probability.  Thus, finding a collision provides a means to distinguish an injective function from a function with a collision.  Therefore, element distinctness is basically a special case of the collision problem for random functions.  This view is in contrast to most existing quantum query results, which treat collision and element distinctness as highly related but distinct problems.

Another compelling answer comes from cryptography: hash functions, one of the central primitives in cryptography, are often modeled as random functions.  This models generic attacks where that do not exploit any particular structure of the hash function and simply evaluate the function on adversarily-chosen inputs.  This model is called the random oracle model~\cite{BR1993}.  

Moreover, the main security requirement for hash functions is that of collision resistance.  In other words, it should be computationally infeasible to find a collision --- two distinct inputs that map to the same output.  Thus, characterizing the quantum query complexity of finding collisions in random functions gives insight into the computational difficulty of finding collisions in hash functions.

Hash functions have wide-spread use in cryptography --- they are used in the construction of efficient signatures, encryption schemes meeting very strong notions of security, and password management, to name a few.  In many of the applications, the domain of the hash function is much, much larger than the codomain, meaning there are many collisions.  Yet, we hope that such collisions are hard to find.  It is important to characterize the difficulty of the collision problem so that we can understand the security of schemes that use hash functions.

\paragraph{Set Equality.}  We also consider the related \emph{set equality} problem, where one is given two injective functions $f$ and $g$ with a common domain and co-domain, with the promise that their ranges are identical or disjoint.  The BHT algorithm can easily be adapted to solve this problem using $O(N^{1/3})$ quantum queries, where $N$ is the size of the domain.  However, the best known lower bound is $\Omega(N^{1/5}/\log N)$ due to Midrijanis~\cite{Midrijanis2004}.  

Improving the lower bound to $\Omega(N^{1/3})$ would have important implications in relating classical and quantum query complexity.  In particular,  Aaronson and Ambainis~\cite{AA2009} show that, for every permutation-symmetric function $f$, the classical randomized query complexity $D(f)$ is at most the ninth power of the quantum query complexity: $D(f)=\tilde{O}(Q(f)^9)$ (where $\tilde{O}$ hides logarithmic factors).  This result uses Midrijanis's lower bound for the set equality problem as a black box.  Improving the lower bound to $\Omega(N^{1/3})$ would give an improved $D(f)=\tilde{O}(Q(f)^7)$ for all permutation-symmetric functions.

\subsection{This work} In this note, we resolve the questions above.  In particular, we prove the following theorem:
\begin{theorem}\label{thm:main} Let $f$ be a random function with domain size $M$ and codomain size $N$.  Assume $M=\Omega(N^{1/2})$.  Then the quantum query complexity of finding a collision with constant probability is $\Theta(N^{1/3})$.\end{theorem}

Note that if $M=o(N^{1/2})$, then there are no collisions with probability approaching 1, so the collision problem becomes meaningless.  Thus, Theorem~\ref{thm:main} completely characterizes the quantum query complexity of the collision problem for all sensible parameters.

We also give an optimal lower bound of $\Omega(N^{1/3})$ for the set equality problem:
\begin{theorem}\label{thm:seteq} Let $f$ and $g$ be random injective functions from $[N]$ to $[M]$ ($M\geq 2N$), conditioned on either:
\begin{itemize}
	\item The ranges of $f$ and $g$ are identical
	\item The ranges of $f$ and $g$ are distinct.
\end{itemize}
Then any algorithm that distinguishes the two cases with bounded error must make $\Theta(N^{1/3})$ quantum queries.\end{theorem}

As discussed above, using Theorem~\ref{thm:seteq} in the proof of Aaronson and Ambainis~\cite{AA2009} gives an improved $D(f)=\tilde{O}(Q(f)^7)$ for all permutation-symmetric functions $f$.  

\paragraph{Techniques} For our lower bound, we start by improving Yuen's lower bound to $\Omega(N^{1/3})$ quantum queries to distinguish a random function from a random permutation.  Our techniques are entirely different, and rely on a lemma from Zhandry~\cite{Zhandry12b} about quantum algorithms distinguishing distributions of functions.  Thus, we can handle the case where $M\leq N$.

We then show how to extend the collision bound to the case where $M=2N$.  Our final step is to decompose a random function on arbitrary domain and codomain as the composition of a sequence of functions, where each component function satisfies $M=2N$.  Using another lemma of Zhandry~\cite{Zhandry12b} about so-called \emph{small-range distributions}, we show that this decomposition is indistinguishable from a random function.  Now, any collision for the overall function must yield a collision for one of the component functions.  We can therefore use the lower bound on the component functions to obtain a lower bound for all functions, as desired.

\medskip

For our upper bound, we use Ambainis's element distinctness algorithm as a black box to give a simple collision finding algorithm.  We note that the BHT algorithm actually works for random functions, provided $M=\Omega(N^{2/3})$.  Using Ambainis's algorithm allows us to cover all $M=\Omega(N^{1/2})$.

\section{The Upper Bound}

\label{sec:upper}

Recall the following theorem of Ambainis~\cite{Ambainis2003}:

\begin{theorem}[\cite{Ambainis2003} Theorem 3] Let $f:[M']\rightarrow[N]$ be a function that has exactly one collision.  Then there is a quantum algorithm making $O((M')^{2/3})$ quantum queries to $f$ that finds the collision with bounded error.
\end{theorem}

We use Ambainis's algorithm as a black box to prove the following:

\begin{theorem}\label{thm:upper} Let $f:[M]\rightarrow[N]$ be a random function, and suppose $M=\Omega(N^{1/2})$.  Then there is a quantum algorithm making $O(N^{1/3})$ quantum queries to $f$ which produces a collision with constant probability.\end{theorem}

The proof is simple, and basically follows the reduction Aaronson and Shi~\cite{AS2004} to prove their lower bound on the element distinctness problem.  Let $M'=N^{1/2}/2$.  Assume for now that $M'\leq M$.  Choose a random subset $S$ of size $M'$.  With probability approaching $1/e$ as $N$ goes to $\infty$, $S$ will contain exactly one collision for $f$.  Let $f'$ be the restriction of $f$ to $S$, and run Aimbanis's algorithm on $f'$.  Then with probability essentially $1/e$, Ambainis's algorithm will return a collision.  The query complexity of Ambainis's algorithm is $O((M')^{2/3})=O(N^{1/3})$, as desired.  In the case where $M>>M'$, this can be repeated multiple times to obtain arbitrarily high success probabilities.

If $M'$ is not significantly smaller than $M$, then there is a constant probability $p$ that no collisions will be found.  However, we can tweak the above algorithm to give us a collision with probability arbitrarily close to $1-p$.

\section{The Lower Bound}

\label{sec:lower}

We now prove our lower bound:
\begin{theorem}\label{thm:lower} There is a universal constant $C$ such that the following holds.  Let $f:[M]\rightarrow[N]$ be a random function.  Then any algorithm making $q$ quantum queries to $f$ outputs a collision for $f$ with probability at most $C(q+1)^3/N$.
\end{theorem}
Thus, to obtain a collision with bounded error requires $q=\Omega(N^{1/3})$ quantum queries.

\paragraph{Tools} We will be using two important tools due to Zhandry~\cite{Zhandry12b}.  The first is the following Lemma: 

\begin{lemma}[\cite{Zhandry12b} Theorem 7.3]\label{lem:quant} There is a universal constant $C'$ such that the following holds.  Let $\Ds_r$ be a family of distributions on functions from $\Xs$ to $\Ys$ indexed by $r\in\mathbb{Z}^+\bigcup\{\infty\}$.  Suppose that, for every integer $k$ and for every $k$ pairs $(x_i,r_i)\in\Xs\times\Ys$, the function $p(r)=\Pr_{f\gets\Ds_{r}}[f(x_i)=r_i\forall i\in\{1,\dots,k\}]$ is a polynomial of degree at most $k$ in $1/r$.  Then any quantum algorithm making $q$ quantum queries can only distinguish the distribution $\Ds_r$ from $\Ds_\infty$ with probability $C' q^3/r$.
\end{lemma}

Zhandry goes on to define the notion of small range distributions, which are essentially the composition of two random functions where the intermediate space is small.  In particular, the small range distribution (for parameter $r$) on functions $f:\Xs\rightarrow\Ys$ is sampled according to the following:
\begin{itemize}
	\item Draw a random function $g$ from $\Xs\rightarrow[r]$.
	\item Draw a random function $h$ from $[r]\rightarrow\Ys$.
	\item Output the composition $f=h\circ g$.
\end{itemize}

Zhandry then uses Lemma~\ref{lem:quant} to prove the following fact about small range distributions:
\begin{lemma}[\cite{Zhandry12b} Corollary 7.5]\label{lem:srdist} Let $f:\Xs\rightarrow\Ys$ be either drawn uniformly at random, or from the small range distribution for parameter $r$.  Then any quantum algorithm making $q$ quantum queries to $f$ can only distinguish the two cases with probability $C' q^3/r$, where $C'$ is the constant from Lemma~\ref{lem:quant}.
\end{lemma}

Using the two lemmas above, we will now prove Theorem~\ref{thm:lower}.

\subsection{The case $\mathbf{M=N}$}

As a first step, we prove the theorem for $M=N$.

We define distributions $\Ds_r$ on functions from $[N]$ to $[N]$, where sampling is obtained by the following process:
\begin{enumerate}
	\item Pick a random function $g:[N]\rightarrow[r]$.
	\item Let $S=\{g(x):x\in[N]\}$.  That is, $S$ is the set of images of $g$.
	\item Pick a random \emph{injective} function $h:S\rightarrow[N]$.
	\item Output the function $f=h\circ g$.
\end{enumerate}

Another way of viewing the distribution $\Ds_r$ is in terms of collision profiles, used in Yuen's original proof~\cite{Yuen2013}.  A collision profile counts, for each $i$, the number of image points of multiplicity $i$.  Then the distribution $\Ds_r$ can be thought of as choosing a collision profile corresponding to a random function from $[N]$ to $[r]$, and then choosing a random function from $[N]$ to $[N]$ with the specified collision profile.

We note three special cases of the distribution $\Ds_r$:
\begin{itemize}
	\item $\Ds_1$.  This distribution just outputs a random constant function.
	\item $\Ds_N$.  This distribution outputs a truly random function from $[N]$ to $[N]$.  Indeed, the function $g$ will be a random function, and the function $h$ can be expanded to a random permutation on $[N]$.  The composition is therefore a random function.
	\item $\Ds_\infty$.  This distribution outputs a random permutation from $[N]$ to $[N]$.  The function $g$ is, with probability 1, an injective function.  Since $h$ is injective, the composition $h\circ g$ is also injective, and therefore a permutation.  Since $h$ is a random injective function, the composition is a random permutation.
\end{itemize}

Therefore, our goal is to show that $q$ quantum queries can only distinguish $\Ds_N$ and $\Ds_\infty$ with probability $C'q^3/N$.  This implies that the probability of finding a collision in $\Ds_N$ is at most $C' (q+2)^3/N$ (we use two extra queries to check if the collision is correct.  The output of this check will can distinguish the two cases).  We need the following lemma:

\begin{lemma}\label{lem:poly} Fix $k$ pairs $(x_i,r_i)\in[N]\times[N]$, and let $p(r)=\Pr_{f\gets \Ds_r}[f(x_i)=r_i\forall i\in\{1,\dots,k\}]$.  Then $p$ is a polynomial in $1/r$ of degree at most $k-1$.
\end{lemma}

Applying Lemma~\ref{lem:quant}, we see that $\Ds_N=\Ds_\Fs$ is indistinguishable from $\Ds_\infty=\Ds_\Ps$ with probability $C'q^3/N$, as desired.

It remains to prove Lemma~\ref{lem:poly}.  Due to the symmetry of the distributions $\Ds_r$, $\Pr_{f\gets \Ds_r}[f(x)=r]=1/N$ for any $(x,r)\in[N]\times[N]$.  Thus, the claim is true for $k=1$.  Now, assume the claim is true for each $k'<k$.  

We can assume without loss of generality that each of the $x_i$ are district.  If not, and $x_i=x_j$, there are two cases:
\begin{itemize}
	\item $r_i=r_j$.  In this case, we can delete the pair $(x_j,r_j)$ since it is redundant. We then invoke Lemma~\ref{lem:poly} using the remaining $k-1$ tuples.
	\item $r_i\neq r_j$.  Then $p(r)=0$ since $f(x_i)$ cannot simultaneously equal $r_i$ and $r_j$.
\end{itemize}

Let $\ell=|\{r_i:i\leq k-1\}|$ be the number of distinct $r_i$ values, not including $r_k$.  Invoking Lemma~\ref{lem:poly} on the first $k-1$ values, we see that $$p'(r)=\Pr_{f\gets \Ds_r}[f(x_i)=r_i\forall i\in\{1,\dots,k-1\}]$$
is a polynomial of degree at most $k-2$ in $1/r$.  We now wish to study the conditional probability $$\Pr_{f\gets \Ds_r}[f(x_k)=r_k:f(x_i)=r_i\forall i\in\{1,\dots,k-1\}]$$

Suppose $r_k=r_i$ for some $i<k$.  This happens exactly when $g(x_k)=g(x_i)$, which happens with probability $1/r$.  Now suppose $r_k\neq r_i$ for any $i<k$.  This means $g(x_k)\neq g(x_i)$ for all $i<k$, which happens with probability $1-\ell/r$.  In this case, $h(g(x_k))=r_k$ with probability $1/(N-\ell)$. Then the probability $f(x_k)=r_k$, conditioned on the other values, is $\frac{(1-\ell/r)}{N-\ell}$.

In either case, the value $$\Pr_{f\gets \Ds_r}[f(x_k)=r_k:f(x_i)=r_i\forall i\in\{1,\dots,k-1\}]$$
is a polynomial of degree 1 in $1/r$.  Combining with $p'(r)$, we see that $p(r)$ is a polynomial of degree at most $k-1$ in $1/r$.  This completes the proof of Lemma~\ref{lem:poly}, and therefore proves Theorem~\ref{thm:lower} for the case $M=N$.

\subsection{The case $\mathbf{M=2 N}$}

Assuming $M=2N$, we can write $[M]$ as $\bin\times [N]$.  Suppose there was an algorithm $A$ that made $q$ queries to a random function $f:\bin\times [N]\rightarrow[N]$, and outputted a collision $x_1,x_2$ with probability $\epsilon$.  We now construct an algorithm $B$ that makes $q$ queries to a random function $g:\bin\times[N]\rightarrow\bin\times[N]$ and outputs a collision with probability $\epsilon/2$.  $B$ works as follows:
\begin{itemize}
	\item Let $f:\bin\times[N]\rightarrow[N]$ be the function obtained from $g$ by dropping the first bit of the output.
	\item $B$ simulates $A$, and when $A$ makes a query to $f$, $B$ forwards the query to $g$, and drops the first bit of the response before returning the response to $A$.  Since quantum operations are reversible, we cannot just disregard the extra bit.  One approach is to uncompute the bit using a second query to $g$.  Another option is to initialize the qubit where the bit will be written to to $\frac{1}{\sqrt{2}}(|0\rangle+|1\rangle)$.  Recall that quantum queries are typically implemented by XORing the response of the query into supplied qubits.  After the query is performed, regardless if the output bit is 0 or 1, the state of the qubit will be $\frac{1}{\sqrt{2}}(|0\rangle+|1\rangle)$.  Thus we can ignore the qubit and throw it away.  
	\item When $A$ outputs a candidate collision $(x_1,x_2)$, $B$ outputs $(x_1,x_2)$ as a collision for $g$.
\end{itemize}

We note that the first bit of output of $g$ is completely independent of $A$'s view.  Also, if $f(x_1)=f(x_2)$, then $g(x_1)=g(x_2)$ exactly when the first bit of output of $g$ is the same in both cases.  This happens with probability $1/2$, independent of $A$'s view.  Thus, if $A$ outputs a collision for $f$ with probability $\epsilon$, $B$ will output a collision for $g$ with probability $\epsilon/2$.

Since $\epsilon/2<C' (q+2)^3/(2N)$, we have that $\epsilon<C' (q+2)^3/N$.

\subsection{The case $\mathbf{M=2^\ell N}$}

Unfortunately, we cannot extend the above analysis to arbitrary $M$, since the probability that the candidate collision is correct drops linearly with $M/N$.  Instead, we play a few tricks to turn an algorithm breaking the $M=2^\ell N$ case into an algorithm breaking the $M=2N$ case.

Suppose we have an adversary $A$ that makes $q$ queries to a function $f:\bin^\ell\times[N]\rightarrow[N]$, and outputs a collision with probability $\epsilon$.

We now define $\ell$ hybrids distributions $\Ds_k$ for $f$ ($k=0,\dots,\ell-1$) as follows:
\begin{itemize}
	\item For $i=1,\dots,k$, let $f_i:\bin^i\times[N]\rightarrow\bin^{i-1}\times[N]$ be a random function for $i=1,\dots,k$.
	\item Let $g_k:\bin^\ell\times[N]\rightarrow\bin^k\times[N]$ be a random function.
	\item Let $f=f_1\circ\dots\circ f_k\circ g$.
\end{itemize}

We observe that the only difference between $\Ds_k$ and $\Ds_{k+1}$ is that $g_k$ is replaced with the composition $f_{k+1}\circ g_{k+1}$.  By the small-range distribution theorem, these two cases are indistinguishable.  In particular, if $A$ can distinguish $\Ds_k$ from $\Ds_{k+1}$ with probability $\delta$, then we can construct an algorithm $B$ that makes $2q$ queries to a random function $g$ that is either $g_k$ or $f_{k+1}\circ g_{k+1}$ and distinguishes the two cases with probability $\delta$.  We note that $B$ must make 2 queries for every query from $A$ in order to uncompute the scratch space used to answer $A$'s queries.  Using Lemma~\ref{lem:srdist}, we know that $\delta < C'(2q+2)^3/2^{k+1}N$ (we've again included the extra two queries needed to check the collision).

Thus, $f$ produces a collision in $\Ds_{\ell-1}$ with probability at least $$\epsilon-\frac{C'(2q+2)^3}{N}\sum_{k=0}^{\ell-1}\frac{1}{2^{k+1}}> \epsilon-\frac{8 C' (q+1)^3}{N}$$

In the case $\Ds_{\ell-1}$, $f$ is generated as $f_1\circ f_2\circ\dots \circ f_\ell$ where $f_i:\bin^i\times[N]\rightarrow\bin^{i-1}\times[N]$ is a random function.  Moreover, any collision for $f$ must yield a collision for one of the $f_i$, which we already know to be difficult.  In particular, if $A$ produces a collision for $f_i$ with probability $\delta$, we know by the $M=2N$ case that $\delta < C' (2q+2)^3)/2^{i-1}N$ (again we need 2 queries for every query $A$ makes in the simulation, and two queries for the check at the end).  

Thus, we have that $$\epsilon-\frac{8C'(q+1)^3}{N}< \frac{C'(2q+2)^3}{N}\sum_{i=1}^\ell\frac{1}{2^{i-1}}$$ meaning $$\epsilon<\frac{24C'(q+1)^3}{N}$$ as desired.

\subsection{General $\mathbf{M}$}

From the above sections, we can conclude that no algorithm can do better than producing a collision with probability $C (q+1)^3/N$, but only for $M/N$ a power of two.  However, it is straightforward to move to general $M$.  In particular, any adversary that produces a collision when the domain is $[M]$ will also work correctly and produce a collision when the domain is $[M']$ for any $M'>M$.  Thus, we can round $M/N$ up to the nearest power of two, and the bound obtained there will be valid.  This completes the proof of Theorem~\ref{thm:lower}.

\section{Lower Bound for Set Equality}

\label{sec:seteq}

In this section, we give an optimal lower bound for the \emph{Set Equality} problem.  In the set equality problem, two injective functions $f$ and $g$ are given with domain $[N]$ and codomain $[M]$ with $M\geq 2N$, with the promise that wither the ranges of $f$ and $g$ are identical, or they are disjoint (notice that the use of $M$ and $N$ is flipped from the above sections).  The goal is to distinguish the two cases.  If the ranges are identical, then the BHT collision finding algorithm can be adapted to find a \emph{claw} of $f$ and $g$: two points $x_1,x_2$ such that $f(x_1)=g(x_2)$.  Thus, using $O(N^{1/3})$ quantum queries, it is possible to distinguish the two cases.  Midrijanis~\cite{Midrijanis2004} shows a $\Omega(N^{1/5}/\log N)$ lower bound for this problem, which is the previous best lower bound.

Our goal is to prove the following theorem, which implies Theorem~\ref{thm:seteq}:

\begin{theorem}\label{thm:seteq2} Let $f$ and $g$ be random injective functions from $[N]$ to $[M]$ ($M\geq 2N$), conditioned on either:
\begin{itemize}
	\item The ranges of $f$ and $g$ are identical
	\item The ranges of $f$ and $g$ are distinct.
\end{itemize}
Then any algorithm that makes a total of $q$ quantum queries to $f$ and $g$ can only solve distinguish the two cases with probability $O(q^3/N)$.\end{theorem}

We now prove Theorem~\ref{thm:seteq}.  Let $r$ be an integer.  Let $f',g'$ be functions from $[N]$ into $[r]$.  Let $S$ be the union of the ranges of $f'$ and $g'$, and let $h$ be a random injective function from $S$ into $[M]$.  We consider 3 cases:

\begin{itemize}
	\item[(1)] $r=N$ and $f'$ and $g'$ are random injective functions.  Then $f$ and $g$ are random injective functions, conditioned on their ranges being identical.  This is the first case of the set equality problem
	\item[(2)] $r=N$ and $f'$ and $g'$ are truly random functions.
	\item[(3)] $r\rightarrow\infty$ and $f'$ and $g'$ are truly random functions.  Then $f'$ and $g'$ are, with probability 1, injective functions and their ranges are distinct.  Therefore, $f$ and $g$ are random injective functions, conditioned on having distinct ranges.  This is therefore the second case of the set equality problem.
\end{itemize}

We now argue that case (1) is indistinguishable from case (2), which is indistinguishable from case (3), which completes the proof.

The only difference between (1) and (2) is that $f'$ and $g'$ are switched from random injective functions to random functions.  Theorem~\ref{thm:lower} shows that the probability of distinguishing these two cases is $O(q^3/N)$.

Now, think of $f'$ and $g'$ as a single function with domain $[N]^2\equiv [2N]$ and range $[r]$.  Similarly, think of $f$ and $g$ as a single function from $[2N]$ to $[M]$.  Then in cases (2) and (3), the combined function $(f',g')$ is just a random function.  Therefore, $(f,g)$ is actually identical to the distribution $\Ds_r$ from Section~\ref{sec:lower}, except with domain $[2N]$ and range $[M]$ where $M\geq 2N$.  Even with the new domain and range, Lemma~\ref{lem:poly} still applies, so we conclude that no quantum algorithm can distinguish the $r=N$ case form the $r=\infty$ case, except with probability $O(q^3/N)$.  Thus, cases (2) and (3) are indistinguishable.

Piecing together, Cases (1) and (3) are indistinguishable, except with probability $O(q^3/N)$, which completes the proof of Theorem~\ref{thm:seteq}.

\section{Conclusion and Open Problems}

Given quantum oracle access to a random function $f:[M]\rightarrow[N]$, we prove that $\Theta(N^{1/3})$ queries are necessary and sufficient to find a collision with bounded error.  One direction for future work would be to look at non-uniform distributions.  For example, what if the outputs of $f$ are not drawn at uniform, but instead drawn from some distribution with (Renyi) entropy $H$.  Does a $\Theta(2^{H/3})$ bound apply?

\bibliographystyle{alpha}
\bibliography{bib}

\end{document}